# Soliton frequency comb generation in a low Q microcavity coupled to a gain microcavity

ZIHAO CHENG[1,2], DONGMEI HUANG[2,3,*], FENG LI[1,2], CHAO LU[1], AND P. K. A. WAI[1,2,4]

[1] Photonics Research Institute, Department of Electronic and Information Engineering, The Hong Kong Polytechnic University, Hong Kong SAR, China
[2] The Hong Kong Polytechnic University Shenzhen Research Institute, Shenzhen 518057, China
[3] Photonics Research Institute, Department of Electrical Engineering, The Hong Kong Polytechnic University, Hong Kong SAR, China
[4] Department of Physics, Hong Kong Baptist University, Hong Kong SAR, China
*Corresponding author: meihk.huang@polyu.edu.hk

Soliton frequency comb generation in coupled nonlinear microcavities is attractive because a coupled microcavity offers more flexibility and possibilities compared to a single nonlinear microcavity. In this paper, we investigate how an amplifying auxiliary cavity affects the bistability region of the main cavity and soliton frequency comb generation. When the auxiliary cavity has a small gain, it can partially compensate for the loss of the main cavity allowing the generation of soliton combs with a relatively low Q–factor in the main cavity. A low Q–factor microcavity would reduce the difficulty of fabrication and extend the microcavity platform to different types of materials. However, if the gain of the auxiliary cavity is too large, a frequency comb cannot be generated because the coupled nonlinear microcavity system is no longer dissipative. Our results provide a theoretical understanding and experimental guidance for the bistability region and soliton frequency comb generation in coupled nonlinear microcavities with an amplifying auxiliary cavity. The results will facilitate the development of chip–scale integrated optical frequency comb sources.

## 1. Introduction

Microcavity based optical frequency combs (microcombs) with compact size, light weight, and low power consumption are promising candidates for comb sources [1]. Their potential for on–chip integration has attracted intense interest from researchers to develop a wide range of applications including atomic spectroscopy [2], coherence optical communications [3], dual–comb spectroscopy [4], LIDAR [5], optical clocks [6], optical coherence tomography [7], radio frequency oscillators [8], and ultrafast optical ranging [9]. To date, the most common microcomb generation scheme is achieved by using a high Q–factor microcavity pumped by a tunable narrow linewidth continuous wave (cw) laser. The soliton frequency comb is excited by tuning the pump laser wavelength from the blue–detuned to the red–detuned side relative to the cavity resonance [10–12]. It remains a challenge to fabricate a low loss microcavity with a high Q–factor. Microcavity losses are mainly due to scattering losses induced by sidewall roughness, especially for microcavities with small free spectral range (FSR), as the relatively large cavity size leads to loss accumulation. In addition, for materials with low intrinsic loss, complex and expensive processing techniques are required to reduce sidewall scattering and produce qualified high Q–factor microcavities [15–17]. Thus, the low loss requirement limits the microcavity sizes and materials for microcomb generation. The generated microcombs with FSR below 10 GHz are still a challenge due to the fabrication bottleneck. In an optical fiber cavity, it is easy to insert a gain element into the cavity to reduce the total cavity loss [13]. However, it is difficult to insert a gain element into a microcavity. Although an on–chip erbium–doped waveguide with gain has recently been reported [14], it is difficult to fabricate a microcavity with different materials due to mode mismatch between waveguides with different materials. In this paper, we propose to solve the problem by coupling an amplifying auxiliary cavity to the main cavity. Compared to mode matching in two on–chip waveguides, it is easier to achieve mode coupling between two microcavities.

Theoretical investigations on the generation of soliton microcombs pumped by a cw laser have already revealed the source of the low loss requirement. The microcomb generation can be modelled by the Lugiato–Lefever equation or the Ikeda map. By performing stability analysis of both models, it was found that the soliton microcomb generation with cw pump depends on the bistability region of the intracavity optical power and the pump power [18–20]. Fig. 1 shows the phase diagram in the parameter space of the cw pump laser power and detuning, where $|F|^2$ is the pump power, $\delta_0 = (\omega_0-\omega_p)T_R/\alpha$ is the pump laser detuning relative to the cavity resonance, $\omega_0$ and $\omega_p$ are the cavity resonance and pump laser frequencies, respectively, and $T_R$ and $\alpha$ are the round–trip time and cavity loss, respectively. The phase diagram shows that when the cavity is pumped by a cw laser, the intracavity optical fields can have different states including chaotic, Turing roll, and soliton states, depending on the pumping conditions. The region available for the generation of soliton microcombs between the solid black curves is the bistability region. The minimum pump laser power for soliton generation is determined by the power at the cusp of the soliton region in Fig. 1. This value is proportional to the cavity loss. If the cavity loss is too high, the minimum pump laser power will exceed a practical value. This is the source of the requirement for low–loss microcavity for soliton microcomb generation.

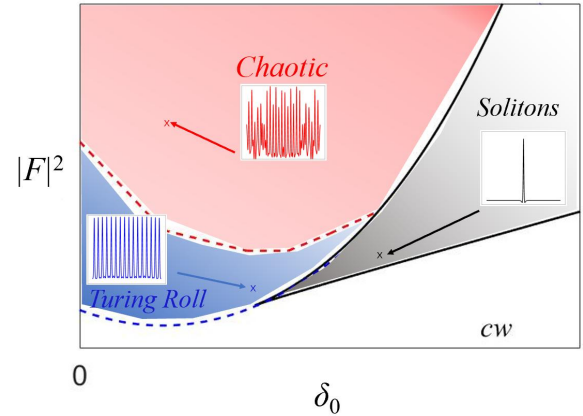

Fig. 1. The phase diagram in the ($\delta_0$, $|F|^2$)–parameter space shows the solution regions of a cw pump microcavity. Different solutions including chaotic states, Turing rolls and solitons can be generated. The soliton region happens to be the bistability region.

In recent years, the coupled microcavities have gradually become attractive especially for the study of parity–time symmetry and exceptional points in photonics [21–23]. Various experimental techniques including gain–loss coupled cavities [22], tuning coupling coefficient [24], and tuning cavity refractive index [25] have been demonstrated to study the parity–time symmetry by adjusting the Q–factor of the coupled microcavities. As for microcomb generation, the coupled microcavity scheme has been investigated for comb generation in normal dispersion microcavity [26], tunable spatiotemporal soliton generation [27], and increasing the pump power utilization efficiency [28]. We have also proposed a novel soliton microcomb generation scheme in coupled nonlinear microcavities by tuning the coupling [29]. From our previous study, the soliton region is tuned by tuning the coupling coefficient. However, the bistability region for soliton microcomb generation in coupled microcavities with a gain auxiliary cavity has never been investigated. In this paper, we will study the soliton microcomb generation in a relatively low Q–factor microcavity coupled to an amplifying microcavity and how the gain affects the bistability region. Our work will provide theoretical guidance for simple soliton microcomb generation and extend the microcomb generation to different materials and platforms. Section II introduces the Ikeda map model for coupled microcavities. How the gain of the auxiliary cavity affects the soliton region of the main cavity will be discussed in Section III. Section IV concludes the paper.

## 2. Coupled microcavity scheme and theoretical model

Figure 2 shows the coupled microcavity scheme being studied. The main cavity is pumped by a cw laser through its waveguide with a coupling coefficient $\theta$. An auxiliary cavity is coupled to the main cavity with coupling coefficient $\theta_c$. The two microcavities are assumed to be identical except that the main cavity is a loss cavity and the auxiliary cavity is a gain cavity.

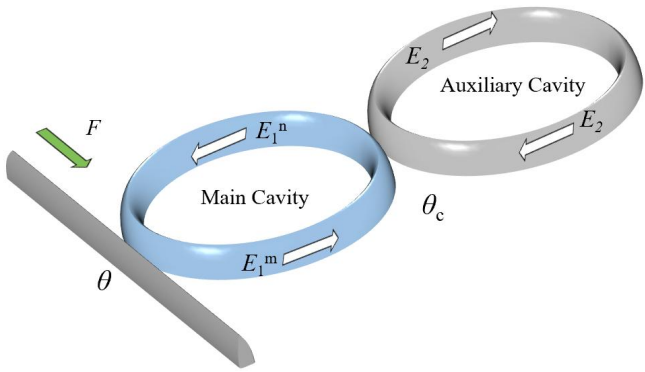

Fig. 2. Coupled microcavity structure under investigation.

The optical fields inside the microcavities are modeled by a set of coupled Ikeda map equations [30],

$$\frac{\partial E_1^{(m,n)}(z,t)}{\partial z} = -i\delta k_0 E_1^{(m,n)} - \frac{\alpha}{2} E_1^{(m,n)} - i\frac{\beta_2}{2}\frac{\partial^2 E_1^{(m,n)}}{\partial t^2} + i\gamma \left|E_1^{(m,n)}\right|^2 E_1^{(m,n)}, \quad (1)$$

$$\frac{\partial E_2(z,t)}{\partial z} = -i\delta k_0 E_2 - \frac{\alpha'}{2} E_2 - i\frac{\beta_2}{2}\frac{\partial^2 E_2}{\partial t^2} + i\gamma \left|E_2\right|^2 E_2, \quad (2)$$

$$E_1^n(0,t) = i\sqrt{\theta_c} E_2(L,t) + \sqrt{1-\theta_c} E_1^m\left(\frac{L}{2},t\right), \quad (3)$$

$$E_1^m(0,t) = i\sqrt{\theta} F + \sqrt{1-\theta} E_1^n\left(\frac{L}{2},t\right), \quad (4)$$

$$E_2(0,t) = i\sqrt{\theta_c} E_1^m\left(\frac{L}{2},t\right) + \sqrt{1-\theta_c} E_2(L,t). \quad (5)$$

where $E_1$ and $E_2$ denote the optical fields inside the main cavity and the auxiliary cavities, respectively. Here m and n denote one half of the main cavity. m denotes the half cavity where the optical field propagates from the waveguide coupling point ($z = 0$) to the cavity coupling point and n denotes the other half cavity with $z = 0$ corresponding to the cavity coupling point as shown in Fig. 2. $\delta k_0$ represents the wave vector of the pump laser in the main cavity and the auxiliary cavities. $\alpha$ and $\alpha'$ are the power losses per unit length of the main and the auxiliary cavities, respectively. $\beta_2$ is the group velocity dispersion and $\gamma$ is the Kerr coefficients of the two cavities. The lengths of the two cavities are $L$, and the amplitude of the cw pump laser is $F$.

In this paper, we use the cusp of the bistability region shown in [29] to evaluate the soliton region. The detuning value and the corresponding pump power at the cusp are presented as $\delta k_p$ and $F_p^2$, respectively. The effect of $\theta_c$ on the bistability region was discussed in our previous report [29], we assume $\theta_c = \theta$ in the following discussions.

## 3. Blue-detuned bistability region influenced by auxiliary cavity with gain

The bistability region of the main cavity is determined numerically by considering the steady–state solutions of Eqs. (1)–(5) [29]. The Q–factor of the main cavity is $7.6\times10^6$, and the cavity parameters are $L$ = 628 μm, $\alpha$ = 19.1 /m, $\beta_2$ = –59 ps$^2$/km, $\gamma$ = 1 /W/m, and $\theta$ = 0.0025. Fig. 3(a) shows the bistability region when the auxiliary cavity has different gain values with $a'=-0.05a$ (dotted curves), $a'=-0.5a$ (dashed curves) and $a'=-a$ (solid curves). The sign of $a'$ is negative because the auxiliary cavity has gain. As the gain of the auxiliary cavity increases, the cusp detuning $\delta k_p$ blue shifts and the cusp power $F_p^2$ decreases. Fig. 3(b) shows the evolution of $\delta k_p$ and $F_p^2$ as $a'$ varies from $-0.01a$ to $-0.5a$. As the gain of the auxiliary cavity increases, $\delta k_p$ blue shifts from $-3.55a$ to $-3.75a$ and $F_p^2$ decreases from 0.3 W to 0.08 W.

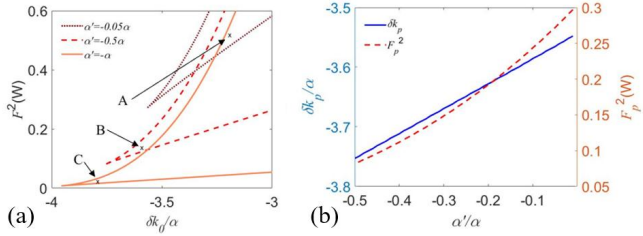

Fig. 3. (a) Bistability regions of coupled microcavities with $a' = -0.05a$, $a' = -0.5a$, and $a' = -a$, the detuning at A, B and C are $-3.2a$, $-3.6a$ and $-3.8a$, respectively, the corresponding pump powers are 0.5, 0.139 and 0.023 W, respectively. (b) Evolution of $\delta k_p$ and $F_p^2$ as a function of the auxiliary cavity gain.

We perform numerical simulations to see if solitons can be generated in the modified bistability regime. We choose two pumping conditions A and B in as shown in Fig. 3(a), where point A is inside the bistability regime of $a' = -0.05a$ and outside the bistability regime of $a' = a$, and point B is inside the bistability regime of $a' = -0.5a$ and outside the bistability regime of $a' = -0.05a$. The detuning values at point A and B are $-3.2a$ and $-3.6a$, respectively, and the corresponding pump powers are 0.5 W and 0.139 W, respectively. The initial conditions in both cavities are weak Gaussian pulses. Figs. 4(a) and 4(b) show the pulse evolution with the pump at point A. When $a' = a$, no pulse can survive, and the initial Gaussian pulse becomes a cw state as shown in Fig. 4(a). When $a' = -0.05a$ in Fig. 4(b), the initial pulse first evolves to a soliton molecule and then becomes a single soliton state. When $a' = -0.05a$, the pump at point B is outside the bistability region and no soliton can form with the initial Gaussian pulse as shown in Fig. 4(c). For a larger gain value, $a' = -0.5a$ and the pump at point B as shown in Fig. 4(d), the initial pulse evolves directly into a soliton molecule state. The simulation results show that the bistability region with an auxiliary gain cavity shown in Fig. 3(a) can support the generation of soliton microcombs.

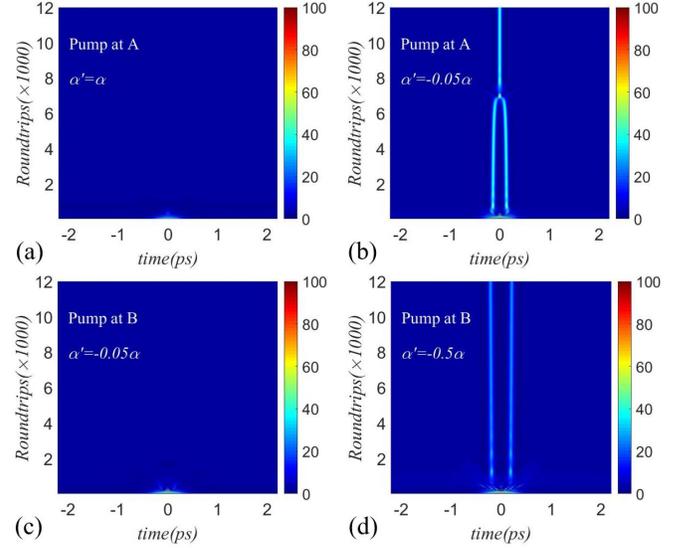

Fig. 4. An initial Gaussian pulse evolution in coupled microcavities with pumping conditions at (a) point A with $a' = a$, (b) point A with $a' = -0.05a$, (c) point B with $a' = -0.05a$, and (d) point B with $a' = -0.5a$.

The theoretical analysis and the simulation results show that if the auxiliary cavity can provide gain, the pump power required to generate the soliton microcomb in the main cavity is reduced compared to a single microcavity. As the gain of the auxiliary cavity increases, the pump power at the cusp of the bistability region in coupled microcavities can be even lower than that in a single cavity with the same parameters as the main cavity. Fig. 5(a) shows the bistability regions of coupled microcavities with an auxiliary gain cavity and a single microcavity. The cavity parameters of the single cavity are the same as the main cavity of the coupled microcavities. The minimum power at the cusp (i.e., the minimum pump laser power for soliton generation) of the bistability region in the coupled cavities with $a' = -0.5a$ is lower than that in a single cavity. Since the power at the cusp of the bistability region is proportional to the cavity loss, Fig. 5(a) confirms that an auxiliary cavity with gain can partially compensate for the loss of the main cavity. This makes it possible to generate soliton microcombs for the low Q–factor microcavities by coupling an auxiliary cavity with gain. The auxiliary cavity can be an on–chip gain cavity or a fiber gain cavity.

It should be noted that when the gain of the auxiliary cavity is too large, i.e., when the coupled cavity system becomes a gain cavity, soliton microcombs cannot be generated. In Fig. 5(b), $\alpha' = -\alpha$ results in a larger bistability region and much lower pump power at the cusp of the bistability region. To see if solitons can be generated under this condition, we choose a pump at point C in the bistability region with $\alpha' = -\alpha$. The corresponding detuning value and pump power are $-3.8\alpha$ and 0.023 W, respectively. We perform a numerical simulation of the pulse evolution at point C with $\alpha' = -\alpha$ and the simulation result is shown in Fig. 5(b). The soliton comb cannot be generated with such a large gain. The generation of soliton microcombs in microcavities requires a balance between the cavity loss and the modulation instability gain, i.e., the cavities must be dissipative. When $\alpha' = -\alpha$, the main cavity becomes an effective gain cavity.

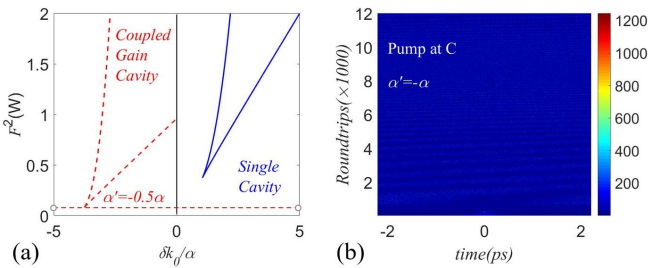

Fig. 5. (a) Bistability region of coupled microcavities with $\alpha' = -0.5\alpha$ (red dashed line) and a single microcavity (blue solid line). (b) An initial Gaussian pulse evolution in coupled microcavities with pumping condition at the point C with $\alpha' = -\alpha$.

## 4. Conclusion

In conclusion, we investigate soliton microcomb generation in a microcavity coupled with an auxiliary cavity with gain. We find that there is a threshold of the auxiliary cavity gain for soliton comb generation. When the gain of the auxiliary cavity is below the threshold, the loss of the main cavity can be compensated by the auxiliary cavity, making it possible to generate soliton frequency combs in low Q microcavities. When the gain of the auxiliary cavity is above the threshold, no soliton is generated because the soliton microcomb is not a stable solution in such a non–dissipative cavity. Our results are important for understanding the generation of soliton frequency combs in coupled microcavities. The idea of using an auxiliary gain cavity to compensate for the loss of the main cavity will open up opportunities for soliton comb generation in microcavities with a wider choice of sizes, material platforms, and simpler fabrication techniques.

**Funding.** National Key R&D Program of China (2020YFB1805901); National Natural Science Foundation of China (62105274); Research Grants Council, University Grants Committee of Hong Kong SAR (PolyU15301022).

**Disclosures.** The authors declare no conflicts of interest.

**Data availability.** Data underlying the results presented in this paper are not publicly available at this time but may be obtained from the authors upon reasonable request.